\documentclass[twocolumn,showpacs,showkeys,preprintnumbers,amsmath,amssymb]{revtex4}

\usepackage{graphicx}

\begin{document}

\title{Random, but not so much\\ 
A parameterization for the returns and correlation matrix of financial time series}
\author{Andr\'e C. R. Martins}
\affiliation{GRIFE - Escola de Artes, Ci\^encias e Humanidades\\
Universidade de S\~ao Paulo, Brazil}

\date{\today}

\begin{abstract}
A parameterization that is a modified version of a previous work is proposed for the returns and correlation matrix of financial time series and its properties are studied. This parameterization allows easy introduction of non-stationarity and it shows several of the characteristics of the true, observed realizations, such as fat tails, volatility clustering, and a spectrum of eigenvalues of the correlation matrix that can be understood as an extension of Random Matrix Theory results. The predicted behavior of this parameterization for the eigenvalues is compared with the eigenvalues of Brazilian assets and it is shown that those predictions fit the data better than Random Matrix Theory.
\end{abstract}

\maketitle

\section{Introduction}

The problem of determining the correct structure of the correlation matrix is an important one in several different applications, and the methods of Random Matrix Theory (RMT)~\cite{wigner,mehta} have been successfully applied to problems in many areas, such as magnetic resonance images~\cite{sengupta}, Meteorology~\cite{santhanam}, and financial time series~\cite{laloux99,plerou99}.

The correct estimation of the correlations in Finance is a fundamental step in portfolio choice~\cite{markowitz}. The observation that most of the eigenvalues of the correlation matrix can be due to noise, therefore, can have important consequences and a model that provides that structure can be a very useful tool in Finance as well as in other areas. RMT does not claim to explain all the eigenvalue spectrum of financial time series, since a few large eigenvalues remain outside its scope. Also, a number of results have been observed that are not in perfect agreement with RMT, such as the observation that noise eigenvalues seem to be a little larger than expected~\cite{kwapien2006} and that correlations can be measured in the supposedly random part of the eigenvalue spectrum~\cite{burda,burdafinancial}. It has also been verified different behaviors of the eigenvalues corresponding to different points of time, suggesting that non-stationary effects might play an important role~\cite{drosdz2000, drosdz2001}.

The role of non-stationarity on the eigenvalue spectrum of the correlation matrix was recently studied and it has been found, by using a model where most eigenvalues are zero in the stationary region, that the non-stationarity can be the cause for the several of the eigenvalues corresponding to the bulk region of the spectrum~\cite{martins2007}. Here, that model will be altered, by introducing random components to the stationary regime. Such an extension will provide a parameterization of the problem where several of the stylized facts about financial series will be observed. Simulations of the model will show that the Mar\u{c}enko-Pastur (MP) distribution ~\cite{marcenkopastur} can be recovered as a limit case for the bulk eigenvalues of the model when more random components are added. The model also allows the introduction of non-bulk, large eigenvalues in the correlation matrix and, therefore, it can be seen as an extension of the results of Random Matrix Theory.

\section{The Model}

In the original model~\cite{martins2007}, the returns $\mu_{i}$ and the correlation matrix $P_{il}$, where both $i=1,\cdots, N$ and $l=1,\cdots, N$ refer to the assets, were obtained from a $N\times M$ matrix $\mathbf \Phi$, that could be a function of the time $t$, $\mathbf \Phi$(t). The matrix $\mathbf \Phi$ components $\varphi_{ij}$, where $i=1,\cdots , N$ represents the different assets and where each value of $j$, $j=1,\cdots , M$, $M\geq3$, can be seen as a collection of $M$ vectors $\mathbf \varphi$, each with $N$ components. Each one of those vectors represents a possible, typical state of the system. Given $\mathbf \Phi$, the average return vector $\mathbf \mu$ and the covariance matrix $\mathbf \Sigma$ and the correlation matrix $\mathbf P$ will be given by
\[
\mu_{i} = E\left[ \varphi_{i} \right]= \frac{1}{M} \sum_{j=1}^{M} \varphi_{ij}
\]
\begin{equation}\label{eq:parametrization}
\Sigma_{il} = \frac{1}{M} \sum_{j=1}^{M} \varphi_{ij}\varphi_{lj} - \mu_{i}\mu_{l},
\end{equation}
\[
P_{il}=\frac{\Sigma_{il}}{\sqrt{\Sigma_{ii}\Sigma_{ll}}}.
\]
\noindent
The observed returns $r_{t}$, at instant $t$, are generated, as usual, by a multivariate normal $N(\mu,\Sigma)$ likelihood.

In this article, a simple, but powerful extension of this model is proposed. Instead of having a matrix  $\mathbf \Phi$ composed of $M\geq3$ vectors, each consisting of parameters to be estimated in order to adjust the model, $\mathbf \Phi$ will be composed of $M+S\geq3$ vectors. The first $M$ vectors play the same role as before~\cite{martins2007}, while we have $S$ new pseudo-parameters, that are actually randomly drawn at each instant of time (even though the $S$ new vectors are not real parameters of the model, since they will be generated randomly, they will be referred to, from now on, as random parameters). With the introduction of the random parameters, all sums in the Equation~\ref{eq:parametrization} are to be performed now on from 1 to $M+S$. This introduces a random element to the model that will cause the return vector and correlation matrix to change in time, even in the stationary case where each of the $\varphi_{ij}$ elements are held constant (at least, for finite values of $S$). In order to preserve the variance associated with each return, the random parameters will follow a normal distribution $N(0,\Sigma_{ii})$ for each asset $i$.

One nice feature of the original model is that, by making each of the components $\varphi_{ij}$ follow a random walk, this generates a non-stationary correlation matrix, with all its properties automatically respected. A simple way to model that is by choosing $\varphi_{ij}(t+1)=\varphi_{ij}(t)+\sigma_\epsilon$. However, for long periods of time, this causes the variance to explode. This is not a problem if one is interested only in the correlation, but, here, the time behavior of the returns will also be investigated. Therefore, a mean-reversion term will be introduced to the random walk, that is
\begin{equation}\label{eq:meanrevrandomwalk}
\varphi_{ij}(t+1)=(1-a)\varphi_{ij}(t)+\sigma_\epsilon,
\end{equation}
where $a$ is a small number that measures the strength of the mean-reversal process ($a=0$ corresponds to no mean-reversal). The effect of this term is negligible for small periods of time as long as $a$ is small enough\footnote{It is interesting to notice that the choices $a$ and $\sigma_\epsilon$ are equivalent to a choice of an average variance for the $\varphi_{ij}$. This can be seen by calculating the variance of Equation~\ref{eq:meanrevrandomwalk} and equating the variances of $\varphi_{ij}$ for $t$ and $t+1$. That point corresponds to the variance value around which the variance of $\varphi_{ij}$ will oscillate.}.

\section{Results}

 \begin{figure}
 \includegraphics[width=0.50\textwidth]{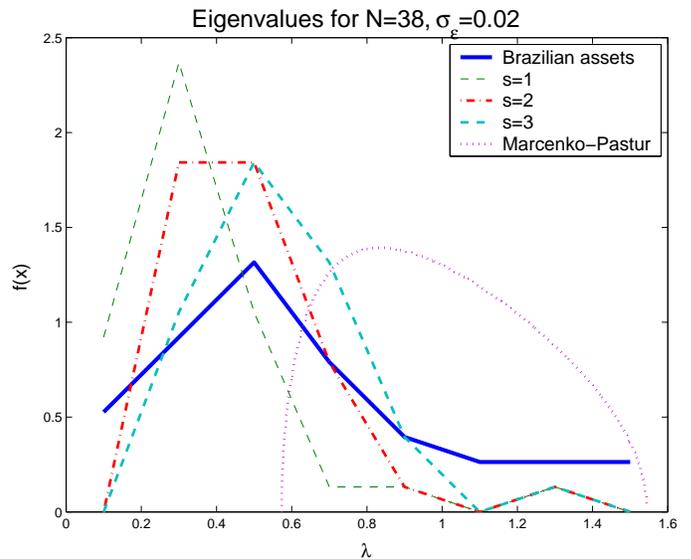}%
 \caption{Observed eigenvalues for $N=38$ Brazilian assets, compared with the simulated results for different values of $s$, in the non-stationary case ($\sigma_\epsilon =0.02$) as well as the MP distribution. \label{Fig:eig38}}
 \end{figure}

Simulations were performed for the proposed model in order to compare it to real data as well as with the Mar\u{c}enko-Pastur distribution~\cite{marcenkopastur}. The real data corresponds to the returns of $N=38$ Brazilian stocks, observed daily from January, 5th, 2004 to July, 28th, 2006, for a total of $T=644$ observations. Figure~\ref{Fig:eig38} shows the behavior of the model for different values of $S$ as a distribution obtained from the histogram of simulated results when $\sigma_\epsilon=0.02$ (the behavior for $\sigma_\epsilon =0.0$ is visually almost identical, with a slightly worse fit, and, therefore, it is not shown here). Notice that the Mar\u{c}enko-Pastur distribution fails to describe the real data, since we are in a finite case, away from the limits where it is expected to be valid. On the other hand, the model here proposed does a much better job, if $S$ is chosen to be 2 or 3. For the simulated results, the two largest eigenvalues are not shown, since they are outside the bulk of random eigenvalues (15.1 and 6, for $S=2$ and 11.9 and 6, for $S=3$). That means that the model not only describes better the observed eigenvalues in the bulk region, but it also generates non-bulk eigenvalues (the real data has one large eigenvalue of 16.2).

\begin{figure}
 \includegraphics[width=0.50\textwidth]{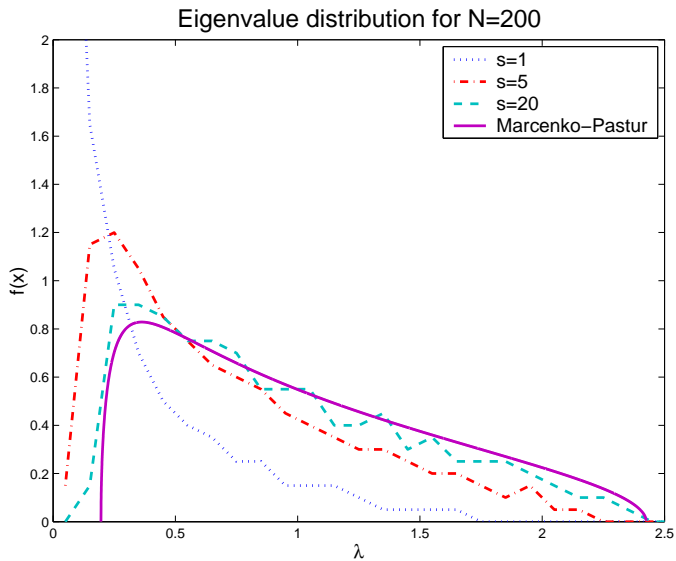}%
 \caption{Simulated eigenvalues for $N=200$, for different values of $s$, in the stationary case ($\sigma_\epsilon =0.0$) compared with the MP distribution.\label{Fig:eig200}}
 \end{figure}

Another interesting feature that can be observed in Figure~\ref{Fig:eig38} is that, as $S$ gets larger, the predicted distribution seems to get closer to the MP distribution. This is actually to be expected. If the $M$  is kept constant, the influence of the real vector parameters in the covariance matrix becomes weaker as $S$ grows. For large $S$, the problem tends to a simple sampling problem and the correlation matrix is obtained from a basically random matrix, therefore the agreement with RMT results. Since $N=38$ is a small number of assets for a good visualization, simulations were run with $N=200$, in order to observe the convergence towards the MP distribution. Those results can be observed in Figure~\ref{Fig:eig200}. 

For $S=0$, the stationary case corresponding to the results shown in Figure~\ref{Fig:eig200}, has only exactly zero eigenvalues, that is, there is only one large peak in the distribution at $\lambda =0$. As $S$ grows, the simulated distributions approaches reasonably fast the MP distribution as can be seen from a reasonable approximation for $S=5$ and an almost exact match when $S=20$. It is also interesting to notice that, although the non-bulk eigenvalues still survive, they are smaller as $S$ grows. This happens because $M$ was kept constant and, therefore, less important for larger values of $S$.

This means that, while $M$ is related to the large eigenvalues, $S$ can be seen as a parameter that measures how close to a random matrix the real data really is, as opposed to a simpler model where only the main eigenvalues exist. In that sense, this model provides an extension of RMT results to cases where the randomization is not complete. It also accounts for the largest observed eigenvalues and, therefore, provides a better fit to real data than RMT.

\begin{figure}
 \includegraphics[width=0.50\textwidth]{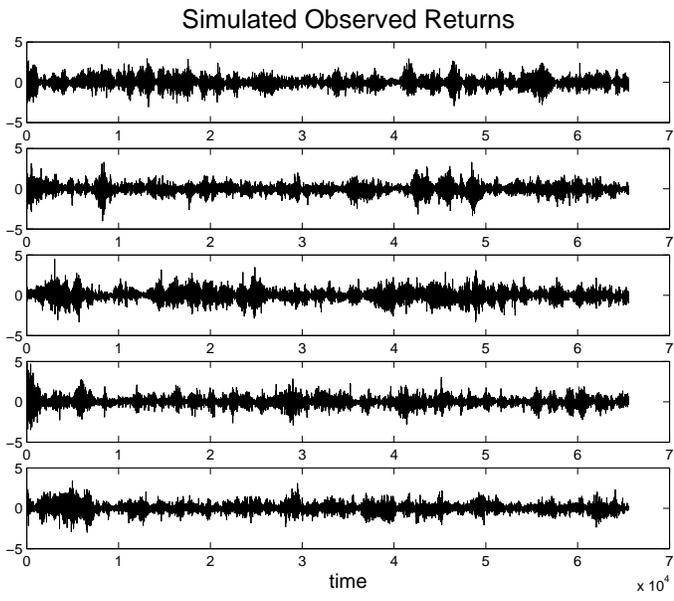}%
 \caption{Simulated returns for $N=5$ assets, with $\sigma_\epsilon =0.02$ and mean reversion given by $a=0.001$.\label{Fig:returns}}
 \end{figure}

Another interesting feature of the simulated time series is the possibility of studying non-stationarity in the covariance matrix and the returns. In order to observe the long run behavior, Equation~\ref{eq:meanrevrandomwalk} was used to generate a mean-reversing random walk in the parameters. Figure~\ref{Fig:returns} show the results for a run with $2^{16}$ time observations of $N=5$ assets, with $M=2$ and $S=1$. The non-stationarity parameters were chosen as $a=0.001$ and $\sigma_\epsilon=0.02$.

It is easy to see the volatility clustering in the time series. Two effects are actually responsible for that; the random walk of the $\varphi_{ij}$ real parameters as well as a less important, but existent effect of the random parameters. That happens because, if the $S$ random parameters are randomly drawn larger than expected once, this will cause the variance at that point in time to increase, making more likely to observe larger random parameters in the next time period.

That is, we have seen that the introduction of random parameters has allowed the proposed model to expand the results of RMT. The resulting model presents a few large eigenvalues (chosen by $M$), a distribution for the bulk eigenvalues that can be made to fit the data better than RMT and made to converge to RMT (by a proper choice of $S$), if necessary, an easy way to introduce non-stationary in returns and in the covariance matrix, and it also shows volatility clustering. Finally, as noted in the original model~\cite{martins2007}, even though normal distributions were used throughout the article, all the observed time series also show an increased kurtosis (except for $\sigma_\epsilon=0$ and $S=0$ or as $S\rightarrow\infty$). This effect diminishes as $S$ grows, since that limit corresponds to a traditional random matrix, but it is important for the smaller values of $S$ that seem to correspond to real problems.


\bibliography{randomparameters}

\end{document}